\documentclass[runningheads]{llncs}
\usepackage{graphicx}
\usepackage{amsmath}

\usepackage[labelfont=bf]{caption}                         
\usepackage[subrefformat=parens]{subfig}
\usepackage{enumitem}  
\usepackage{amssymb}
\usepackage[misc]{ifsym}
\usepackage{graphicx}
\usepackage{threeparttable}
\usepackage{ntheorem}
\usepackage{color}
\usepackage{colortbl}
\usepackage{multirow}
\usepackage{amsmath}
\usepackage{comment}
\usepackage{bm}                                            
\usepackage{etoolbox}                                      
\usepackage{url}
\usepackage{nth}
\usepackage{cite}
\usepackage{balance}
\usepackage[bookmarks=false]{hyperref}                     %
\usepackage{courier}                                       
\usepackage{listings}                                      
\usepackage[]{algpseudocode}
\usepackage{makecell}  
\algrenewcommand\textproc{\texttt}
\makeatletter\let\c@float@type\relax\makeatother
\makeatletter\let\float@addtolists\relax\makeatother
\usepackage{algorithm}
\hypersetup{
    colorlinks = true,
    citecolor  = blue,
    linkcolor  = blue,
    urlcolor   = blue,
}

\makeatletter
\newcommand{\thickhline}{%
	\noalign {\ifnum 0=`}\fi \hrule height 1pt
	\futurelet \reserved@a \@xhline
}

\graphicspath{{./figs/}}
\DeclareMathOperator*{\argmin}{arg\,min}
\newcommand{\cartprod}{\mathbin{\mkern-2mu\times\mkern-2mu}}
\begin{document}

\title{VoxelOpt: Voxel-Adaptive Message Passing for Discrete Optimization in Deformable Abdominal CT Registration}

\author{
Hang Zhang \inst{1} \and 
Yuxi Zhang \inst{2} \and 
Jiazheng Wang \inst{2} \and 
Xiang Chen \inst{2} \and 
Renjiu Hu \inst{1}  \and 
Xin Tian \inst{3} \and 
Gaolei Li \inst{4} \and 
Min Liu  \inst{2} \textsuperscript{(\Letter)} 
}
\authorrunning{H. Zhang \emph{et al}.}

\institute{Cornell University, Ithaca, USA \and
Hunan University, Changsha, China \\ \email{liu\_min@hnu.edu.cn} \and
University of Oxford, Oxford, UK  \and 
Shanghai Jiao Tong University, Shanghai, China
}

\maketitle

\begin{abstract}

Recent developments in neural networks have improved deformable image registration (DIR) by amortizing iterative optimization, enabling fast and accurate DIR results.  
However, learning-based methods often face challenges with limited training data, large deformations, and tend to underperform compared to iterative approaches when label supervision is unavailable.
While iterative methods can achieve higher accuracy in such scenarios, they are considerably slower than learning-based methods. 
To address these limitations, we propose \textbf{VoxelOpt}, a discrete optimization-based DIR framework that combines the strengths of learning-based and iterative methods to achieve a better balance between registration accuracy and runtime.
VoxelOpt uses displacement entropy from local cost volumes to measure displacement signal strength at each voxel, which differs from earlier approaches in three key aspects.  
First, it introduces voxel-wise adaptive message passing, where voxels with lower entropy receives less influence from their neighbors.  
Second, it employs a multi-level image pyramid with 27-neighbor cost volumes at each level, avoiding exponential complexity growth.  
Third, it replaces hand-crafted features or contrastive learning with a pretrained foundational segmentation model for feature extraction.  
In abdominal CT registration, these changes allow VoxelOpt to outperform leading iterative in both efficiency and accuracy, while matching state-of-the-art learning-based methods trained with label supervision.
The source code will be available at \url{https://github.com/tinymilky/VoxelOpt}.


\end{abstract}

\section{Introduction}
\label{sec:introduction}

Traditional deformable image registration (DIR) typically minimizes a Horn-Schunck-type energy function \cite{horn1981determining}, combining a dissimilarity metric and a smoothness regularizer.  
While effective, these methods \cite{avants2008symmetric,heinrich2013mrf} rely on computationally expensive iterative optimization, resulting in slow runtime for large volumetric data. 
In contrast, learning-based methods, pioneered by VoxelMorph \cite{balakrishnan2019voxelmorph}, amortize iterative optimization by training neural networks to predict deformations in a single forward pass.  
This enables fast image registration and can potentially achieve higher accuracy when trained with label supervision.  
However, learning-based approaches struggle with limited training data, large deformations, and may underperform iterative methods in the absence of supervision.  

To address these limitations, recent methods \cite{wu2022nodeo,van2023robust,siebert2024convexadam} have combined learning-based efficiency with iterative optimization.  
For example, ConvexAdam \cite{siebert2024convexadam} performs iterative optimization within deep learning frameworks (e.g., PyTorch), leveraging parallel GPU processing to achieve considerable speed improvements over classical approaches.  
Additionally, lightweight neural networks can be integrated into the optimization loop \cite{wu2022nodeo,van2023robust} to enhance accuracy.  
However, two critical challenges remain:  
1) These hybrid methods \textbf{remain slower than pure learning-based approaches} despite their acceleration over classical methods;  
2) They rely solely on photometric dissimilarity measures, which are empirically shown to provide \textbf{limited anatomical correspondence information} \cite{jena2024deep}.  

We hypothesize that message passing mechanisms \cite{krahenbuhl2011efficient} have been largely overlooked in existing registration methods.  
As shown in the original Horn-Schunck variational formulation \cite{horn1981determining}, displacements in uniform or smooth regions with weak intensity gradients must be inferred from boundaries or gradient-rich regions.  
This "filling" process, achieved through an isotropic diffusive regularizer, is essentially a form of message passing.  
To our knowledge, all leading registration methods, both learning-based and iterative, rely on this isotropic regularizer for message passing.
However, this serves as one of the major factors limiting the efficiency of iterative methods.  
Propagating displacement signals requires numerous iterations, resulting in slow convergence, while stronger regularization, though accelerating convergence, inevitably oversmooths the displacement field.  

To address these challenges, we propose \textbf{VoxelOpt}, a voxel-adaptive message passing framework for discrete optimization in deformable image registration.  
\textbf{First}, VoxelOpt quantifies displacement signal strength using the entropy of a probabilistic 27-neighbor cost volume, guiding adaptive message passing: voxels with strong signals (low entropy) retain their information, while those with weak signals (high entropy) receive more influence from neighbors.  
This resembles spatially adaptive filtering \cite{zhang2023spatially}.
\textbf{Second}, while discrete optimization-based DIR methods \cite{steinbrucker2009large,heinrich2013mrf,heinrich2019closing,siebert2024convexadam} typically converge faster than gradient descent-based approaches, they suffer from exponentially growing displacement search spaces.  
In contrast, VoxelOpt employs a multi-level image pyramid, where the cost volume at each level considers only locations within a single voxel (i.e., the 27-neighborhood).  
This design avoids exponential space growth while remaining compatible with the large-deformation diffeomorphic framework \cite{beg2005computing}.  
\textbf{Third}, inspired by prior work \cite{chen2024textscf,zhang2024memwarp,zhang2024slicer} demonstrating that segmentation-derived feature maps can enhance registration, we adopt a pretrained foundational CT segmentation model trained on large-scale datasets for feature extraction.  
With foundation models becoming increasingly accessible, this avoids complex contrastive learning while providing richer semantic context than raw images \cite{jena2024deep}.

Preliminary results on an abdominal CT dataset with limited training data and large deformations demonstrate the effectiveness and efficiency of VoxelOpt.  
It outperforms the best unsupervised learning method by 14.7\% in Dice and the best iterative method by 9.2\%, with substantial runtime reduction.  
VoxelOpt also matches the performance of the best semi-supervised learning method without requiring label supervision.  

\section{Methodology}
\label{sec:method}
\subsection{Preliminaries}

Deformable image registration (DIR) \cite{sotiras2013deformable,haskins2020deep,viergever2016survey} is typically formulated as a variational optimization problem:  
\begin{equation}  
    \hat{\mathbf{u}} = \argmin_{\mathbf{u}} \left\{ d_s(\textbf{\textit{f}}, \textbf{\textit{m}} \circ (\mathbf{u} + \mathbf{I}_d)) + \lambda r(\mathbf{u}) \right\}.  
    \label{eq:var_opt}  
\end{equation}  
Here, \(\textbf{\textit{f}}\), \(\textbf{\textit{m}}\), \(\mathbf{I}_d\), and \(\mathbf{u}\) denote the fixed image, moving image, identity transformation grid, and displacement field, respectively.  
The terms \(d_s(\cdot, \cdot)\) and \(r(\cdot)\) represent the dissimilarity function and smoothness regularizer, while \(\lambda\) controls the regularization strength. 
For discrete optimization in image registration, we follow prior work \cite{heinrich2014non,heinrich2019closing,zhang2025unsupervised} and construct the energy function using local cost aggregation with quadratic relaxation \cite{chambolle2004algorithm,zach2007duality}, formulated as follows:  
\begin{align}
    \hat{\mathbf{u}},\hat{\mathbf{v}}=\argmin_{\mathbf{u},\mathbf{v}} ~~&  
    \sum^{\mathbf{\Omega}}_x\mathbf{C}^k(x)\circ\mathbf{v}(x) + \frac{1}{2\theta} \|\mathbf{v} - \mathbf{u}\|^2 + \lambda r(\mathbf{u}), \nonumber
    \\
\text{s.t.}  ~~& \forall ~x \in \mathbf{\Omega}, \mathbf{v}(x)\in \mathcal{L}_k . 
    \label{eq:discrete_reg}
\end{align}
Here, $\mathbf{\Omega} \subseteq\mathbb{Z}^3$ is the spatial space of the image,  \(\frac{1}{2\theta}\) is the interaction term coefficient, \(n = h \cartprod w \cartprod d\) is the image spatial size, $\mathcal{L}_k = \{0, \pm 1, \pm 2, \ldots, \pm k\}^3 $.  
The intensity at position \(x\) is \(\textbf{\textit{f}}(x)\) and \(\textbf{\textit{m}}(x)\), with displacement \(\mathbf{v}(x) \in \mathbb{R}^{3}\) and \(\mathbf{u}(x) \in \mathbb{R}^{3}\).  
The cost volume \(\mathbf{C}^k\) is a 6D tensor: the first three dimensions represent spatial coordinates, and the last three define the local cost volume of kernel size \(k\) at each voxel.  
We denote the local 3D cost volume at position \(x\) as \(\mathbf{C}^k(x) \in \mathbb{R}^{(2k+1) \cartprod (2k+1) \cartprod (2k+1)}\), pre-computed by measuring dissimilarities between each voxel and its neighbors:   
\begin{equation}
    \mathbf{C}^k(x,o) = d_s\big(\textbf{\textit{f}}(x), \textbf{\textit{m}}(x+o)\big).
    \label{eq:cost_volume}
\end{equation}
Here, \(o \in \mathcal{L}_k\) enumerates the offsets in a local neighborhood of kernel size \(k\).  

\subsection{Voxel-wise Displacement Entropy} 
\label{sec:entropy}
To enable voxel-adaptive discrete optimization, we quantify the signal strength at each voxel location.  
This strength reflects the confidence of the derived displacement vector and determines the extent of information exchange with neighboring voxels.  
Specifically, voxels with stronger signals should prevent their information from being diluted, while those with weak signals should put higher weights on receiving information from neighboring regions. 

Under sufficiently downsampled input images, we can assume the largest displacement magnitude is within one voxel.  
Thus, we compute the displacement entropy \(\mathbf{E} \in \mathbb{R}^{h\cartprod w \cartprod d}\) using the 6D probabilistic cost volume tensor \(\mathbf{P}^1\) as follows:  
\begin{equation}
    \mathbf{P}^1(x,o) = \frac{\text{exp}(-\mathbf{C}^1(x,o)/\beta)}{\sum_p^{\mathcal{L}_1}\text{exp}(-\mathbf{C}^1(x,p)/\beta)},~~
    \mathbf{E}(x) = \sum_o^{\mathcal{L}_1} -\mathbf{P}^1(x,o)\text{log}(\mathbf{P}^1(x,o)),
    \label{eq:entropy}
\end{equation}
where \(\exp(\cdot)\) is the exponential operator, \(\log(\cdot)\) is the logarithm operator, and \(\beta\) controls the temperature.
We refer to \(\mathbf{E}(x)\) as the per-voxel displacement entropy.  

\textbf{Displacement Entropy Interpretation:}  
The per-voxel entropy \(\mathbf{E}(x)\) quantifies the uncertainty of the displacement derived from the cost volume (by taking the argmin).
For voxels in uniform or smooth regions, entropy tends to be high due to similar neighboring intensities, resulting in a uniform probability distribution.  
In contrast, voxels near boundaries or in texture-rich regions exhibit sparse probability distributions, leading to lower entropy.  
Thus, smaller \(\mathbf{E}(x)\) indicates stronger displacement signals with less uncertainty, and vice versa.

\begin{figure}[!t]
    \centering
    \includegraphics[width=0.85\columnwidth]{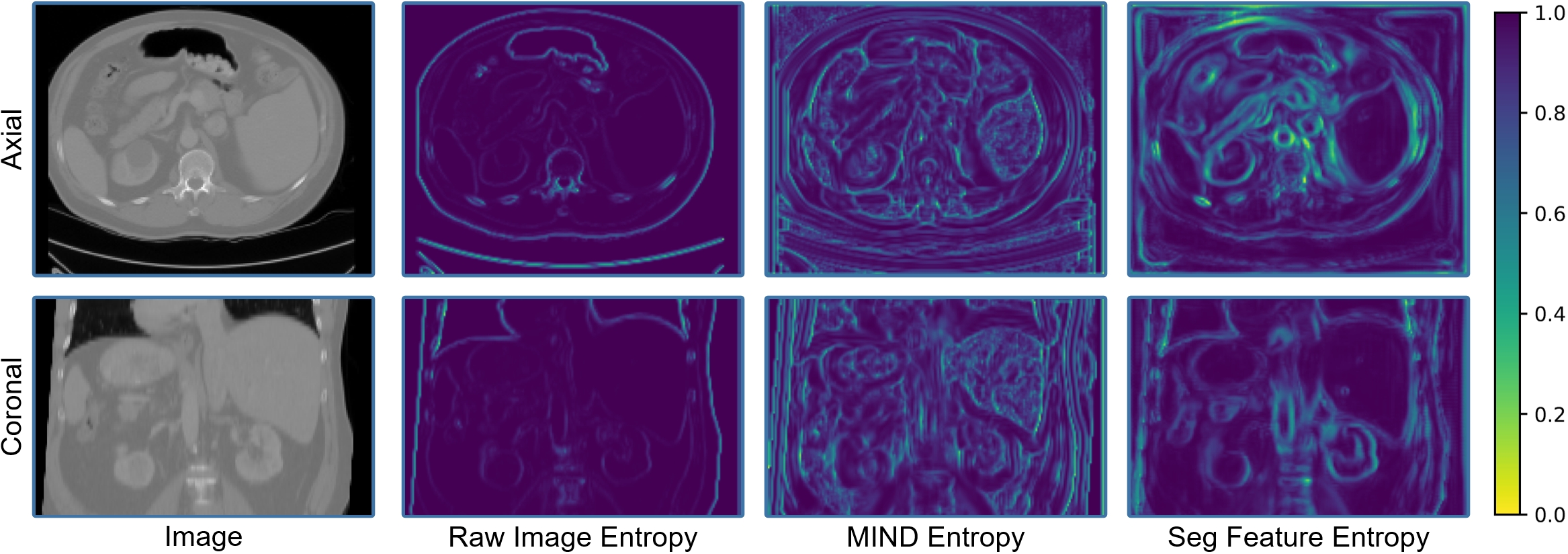}
    \caption{  
    Visual examples of how features affect entropy distribution.  
    The fixed image is shown in the first column; the moving image is manually shifted by one voxel.  
    Cost volumes are computed using a kernel size of 1, and per-voxel entropies are derived via Eq.~\eqref{eq:entropy}, then normalized via Eq.~\eqref{eq:norm_scale}.
    }
    \label{fig:entropy_distribution}
\end{figure}

\subsection{Feature Impact on Entropy Distribution}  
Lower entropy values across more voxels simplify the overall displacement field extraction, as displacement signals propagate faster and more reliably.  
Thus, image features should promote a wider distribution of voxels with lower entropies.  
However, while certain features yield better entropy distributions, low entropy can also arise from local noise.  
To evaluate feature impact on entropy, we study raw image intensity, MIND feature maps \cite{heinrich2012mind}, and feature maps from a pretrained foundational segmentation model \cite{liu2023clip} on abdominal CT scans.

As shown in Fig.~\ref{fig:entropy_distribution}, both MIND and segmentation (seg) feature maps produce more widely distributed low-entropy voxels compared to raw images.  
While raw images exhibit high entropy even at boundaries of large objects like the liver, seg features accurately capture boundary signals without introducing local noise in uniform regions.  
In contrast, MIND's self-similarity introduces considerable amount of local noise in uniform areas such as the liver. 

\subsection{Voxel-wise Adaptive Message Passing}
To solve Eq.~\eqref{eq:discrete_reg}, we employ coordinate descent, alternating between \(\mathbf{v}\) and \(\mathbf{u}\) while progressively reducing \(\theta\) to finally achieve \(\hat{\mathbf{v}} \approx \hat{\mathbf{u}}\).
Then the optimization of Eq.~\eqref{eq:discrete_reg} can be decomposed into two subproblems as follows:
\begin{subequations} \label{eq:unrolled_sp} 
\begin{align} 
\hat{\mathbf{v}} = \argmin_{\mathbf{v}} &~ \frac{1}{2\theta} \|\mathbf{v} - \hat{\mathbf{u}}\|^2 + \sum^{\mathbf{\Omega}}_x\mathbf{C}^k(x)\circ\mathbf{v}(x),  ~\text{s.t.} ~\forall ~x \in \mathbf{\Omega}, \mathbf{v}(x)\in \mathcal{L}_k; \nonumber \\ 
\hat{\mathbf{u}} = \argmin_{\mathbf{u}} &~ \frac{1}{2\theta} \|\hat{\mathbf{v}} - \mathbf{u}\|^2 + \lambda r(\mathbf{u}). \tag{5}
\end{align} 
\end{subequations}
Here, each subproblem is convex and can be solved globally \cite{steinbrucker2009large}.  
The \(\mathbf{v}\) subproblem is solved optimally and pointwise across \(x \in \mathbf{\Omega}\), as each voxel's displacement is independent of others.  
Notably, since \(\mathbf{v}(x)\) is restricted to a discrete space, we approximate the optimizer by propagating the interaction term's effects to the cost volume and performing an argmin search.  
When $r(\mathbf{u})=\|\nabla\mathbf{u}\|$, typically the \(\mathbf{u}\) subproblem can be solved through the fixed-point iteration \cite{chambolle2004algorithm}, yielding \(\mathbf{u}\) as a smoothed version of \(\hat{\mathbf{v}}\). 
This can be approximated by applying Gaussian filtering to \(\hat{\mathbf{v}}\) as \(\hat{\mathbf{u}}=\mathcal{K} * \hat{\mathbf{v}}\) across spatial locations.

\textbf{Adaptive Message Passing:}
Solving the \(\mathbf{u}\) subproblem resembles mean-field inference, where displacement signals act as messages passed between neighboring voxels.  
While widely used in prior approaches \cite{heinrich2014non,siebert2024convexadam}, isotropic Gaussian filtering can dilute strong signals with weak ones, leading to oversmoothing.  
This may require larger offset search spaces, increasing complexity exponentially.  

To address this issue, we propose voxel-wise adaptive message passing.  
Using voxel-wise displacement entropy (see \S\ref{sec:entropy}), we determine signal strength, which controls the extent of information each voxel receives from its neighbors: stronger signals receive less input, while weaker signals receive more.  
Specifically, we use normalized voxel entropies in \((0,1)\) to compute a per-voxel \(\sigma(x)\), adaptively controlling the information received during each message passing phase:  
\begin{equation}  
    \sigma(x) = \alpha \log\left(\frac{\mathbf{E}(x)}{\max(\mathbf{E})} + 1\right),  
    \label{eq:norm_scale}
\end{equation}  
where \(\max(\cdot)\) computes the tensor's maximum value, \(\log(\cdot)\) is the logarithm operator, and \(\alpha\) controls the overall blurring strength.  

\begin{figure}[!t]
    \centering
    \includegraphics[width=1.0\columnwidth]{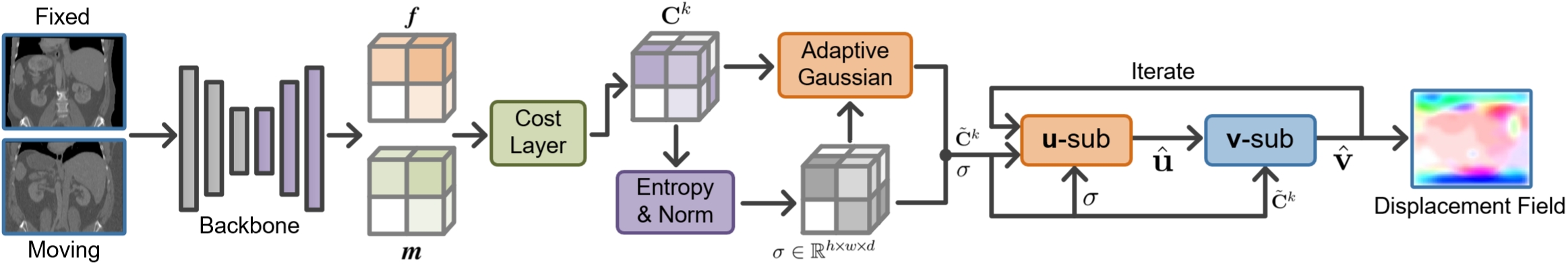}
    \caption{  
    Schematic of the VoxelOpt framework.
    }  
    \label{fig:voxelopt_framework}
\end{figure}

\subsection{Overall Framework of VoxelOpt}
Unlike iterative methods requiring many iterations to converge, VoxelOpt uses discrete optimization \cite{heinrich2014non,siebert2024convexadam} and converges in just 6 iterations.  
Crucially, VoxelOpt operates like a pre-trained learning-based method without training, as it also computes the displacement field in a single forward pass.  

As shown in Fig.~\ref{fig:voxelopt_framework}, fixed and moving images are processed by a pretrained foundational network \cite{liu2023clip} to extract feature maps \(\textbf{\textit{f}}\) and \(\textbf{\textit{m}}\), which are fed into a cost layer (Eq.~\eqref{eq:cost_volume}) to compute the 6D cost volume \(\mathbf{C}^k\).  
Using Eq.~\eqref{eq:entropy}, \(\mathbf{C}^k\) computes voxel-wise displacement entropy, normalized and scaled via Eq.~\eqref{eq:norm_scale} to obtain \(\sigma \in \mathbb{R}^{h \cartprod w \cartprod d}\), guiding adaptive message passing.  
The voxel-adaptive \(\sigma\) filters \(\mathbf{C}^k\) spatially to produce \(\tilde{\mathbf{C}}^k\), which, along with \(\sigma\), supports iterative optimization of the \(\mathbf{v}\)- and \(\mathbf{u}\)-subproblems, yielding the final displacement field.  

With kernel size \(k = 1\) (capturing displacements within one voxel using 27 neighbors), we employ an \(N\)-level Laplacian image pyramid for large-deformation diffeomorphic transformation.  
After extracting feature maps \(\textbf{\textit{f}}\) and \(\textbf{\textit{m}}\), we trilinearly downsample them \(2\cartprod\) at each level to construct the \(N\)-level pyramid.    
At each level, the fixed feature map and the moving feature map (warped by the previous level's deformation field) undergo discrete optimization to compute the residual field.  
Each residual field is processed via scaling and squaring \cite{ashburner2007fast}, and the final field is obtained by composing all levels' fields.  

\section{Experiments \& Results}

\subsection{Datasets}  
We evaluated our method on a public dataset of 30 abdominal CT scans \cite{xu2016evaluation}, each with segmentation masks for 13 organs.  
The dataset was split into 380 training pairs (20×19), 6 validation pairs (3×2), and 42 testing pairs (7×6) to compare with both iterative and learning-based methods.  
Images were first resampled to 2 mm voxel size, resized to 192×160×256, and min-max normalized to \([0,1]\) after clipping intensities to \([-800,500]\).
In the unsupervised setting, no segmentation masks were used; in the semi-supervised setting, masks with 13 organ labels were used to compute Dice loss measuring anatomical alignment only during training.

\subsection{Baselines, Implementations \& Evaluation Metrics}
\label{sec:base_imp_eva}
\textbf{Baseline Methods:}
We compare VoxelOpt against three categories: iterative optimization, unsupervised learning-based, and semi-supervised learning-based methods.  
Iterative methods include Ants SyN \cite{avants2008symmetric}, Deeds \cite{heinrich2013mrf}, ConvexAdam \cite{siebert2024convexadam}, NODEO \cite{wu2022nodeo}, ccINR \cite{van2023robust}, and PRIVATE \cite{hu2024a}.  
Learning-based methods include VoxelMorph \cite{balakrishnan2019voxelmorph}, FourierNet \cite{jia2023fourier}, RDP \cite{wang2024rdp}, and CorrMLP \cite{meng2024correlation}.  

\textbf{Baseline Implementation Details:}
All learning methods were implemented using the CorrMLP framework and trained under identical conditions: 100 epochs on an A6000 GPU with PyTorch, Adam optimizer (learning rate \(1 \times 10^{-4}\)), batch size 1, regularization \(\lambda = 1\), and NCC dissimilarity (window size 9).  
Scaling and squaring \cite{arsigny2006log} with 7 integration steps ensured diffeomorphic transformations, as in \cite{balakrishnan2019voxelmorph}.  
Iterative methods were evaluated using their public code repositories, with configurations tuned for optimal performance (see remarks in results).

\begin{table*}[t]  
\centering  
\caption{  
Quantitative results on the abdominal dataset, with metrics averaged over the 42 testing pairs.  
``Initial'' refers to baseline results before registration, and a down arrow ($\downarrow$) indicates that lower values are better.  
The best-performing metric in each column is bolded, and the second-best is underlined.  
Each method within the respective category is sorted by Dice score, from low to high.  
}  
\resizebox{0.99\columnwidth}{!}{  
\begin{tabular}{rclcccr}  
    \hline  
    \hline  
    Model & Type & Remarks & Dice (\%) & HD95 $\downarrow$ & SDLogJ $\downarrow$ & Runtime (s) $\downarrow$ \\  
    \hline  
    Initial & - & - & 30.86 & 29.77 & - & - \\  
    \hline
    Ants SyN \cite{avants2008symmetric} & \multirow{6}{*}{Iterative Methods} & Affine+Deform & 37.02 & 25.48 & \textbf{0.07} & 22.6  \\
    ccINR \cite{van2023robust} &  & 2500 iterations & 39.07 & 24.09 & 0.21 & 77.1 \\  
    PRIVATE \cite{hu2024a} & & 500 iterations & 45.49 & 24.11 & 0.13 & 7.1 \\  
    ConvexAdam \cite{siebert2024convexadam} & & 80 iterations & 50.23 & 22.60  & 0.13  & 7.0 \\  
    NODEO \cite{wu2022nodeo} & & 300 iterations & 51.75 & 22.60 & 0.15 & 18.5 \\  
    Deeds \cite{heinrich2013mrf} & & MIND+NCC & 53.57 & 20.08 & \underline{0.12} & 110.1 \\  
    \hline  
    FourierNet \cite{jia2023fourier} & \multirow{4}{*}{\makecell{Learning-based Methods \\ Unsupervised} } & Start channels 32 & 41.83 & 25.25 & \textbf{0.11} & $<1.0$ \\  
    VoxelMorph \cite{balakrishnan2019voxelmorph} &  & Start channels 32 & 41.90 & 25.97 & 0.12 & $<1.0$ \\  
    RDP \cite{wang2024rdp} & & Channels 16 & 50.91 & 22.88 & 0.14 & $<1.0$ \\  
    CorrMLP \cite{meng2024correlation} & & Enc 8, Dec 16 & 51.01 & 22.80 & 0.13 & $<1.0$ \\  
    \hline  
    FourierNet \cite{jia2023fourier} & \multirow{4}{*}{\makecell{Learning-based Methods \\ Semi-supervised} } & Start channels 32 & 42.80 & 22.95 & 0.13 & $<1.0$ \\  
    VoxelMorph \cite{balakrishnan2019voxelmorph} &  & Start channels 32 & 47.05 & 23.08 & 0.13 & $<1.0$ \\  
    CorrMLP \cite{meng2024correlation} & & Enc 8, Dec 16 & 56.58 & 20.40 & 0.16 & $<1.0$ \\  
    RDP \cite{wang2024rdp} &  & Channels 16 & \textbf{58.77} & \underline{20.07} & 0.22 & $<1.0$ \\  
    \hline  
    \textbf{VoxelOpt (ours)} & \multirow{1}{*}{Iterative Methods} & 6 iterations, $k=1$ & \underline{58.51} & \textbf{18.54} & 0.21 & $<1.0$ \\  
    \hline  
    \hline  
\end{tabular}  
\label{tab:abct}  
}  
\end{table*}  

\textbf{VoxelOpt Implementation Details:}  
To handle large deformations, we employ a 5-level image pyramid with a local cost volume kernel size of \(k = 1\) (unless specified otherwise).  
The blurring strength cap in Eq.~\eqref{eq:norm_scale} is set to \(\alpha = 1.5\).  
Following prior work \cite{heinrich2014non}, we use 6 iterations with gradually decaying \(\theta = \{150, 50, 15, 5, 1.5, 0.5\}\) in solving Eq.~\eqref{eq:unrolled_sp}.  
VoxelOpt adopts the \(L_1\) dissimilarity function and scaling-and-squaring with 7 integration steps for diffeomorphic transformations.
The adaptive 3D Gaussian filtering is implemented as a sequence of three 1D Gaussian filters applied along the x-, y-, and z-axes, leveraging the separability of the Gaussian kernel.
The backbone network uses the foundational segmentation model \cite{liu2023clip}, with pre-softmax feature maps as inputs.  
All experiments were conducted on the same computing machine and environment as baseline methods.

\textbf{Evaluation Metrics:}
Anatomical alignment is evaluated using Dice Similarity Coefficient (Dice) and 95\% Hausdorff Distance (HD95), while smoothness is assessed via the standard deviation of the log Jacobian determinant (SDlogJ).  
Runtime, averaged over 42 testing pairs, is denoted as \(<1\) for sub-second durations, as sub-second differences for volumetric image registration are negligible.

\subsection{Results \& Analysis}

Quantitative results of VoxelOpt compared to baseline methods are shown in Table~\ref{tab:abct}.  
Deeds \cite{heinrich2013mrf}, using Markov random fields and tree-DP for discrete optimization, achieves the highest registration accuracy among iterative methods but requires the most runtime.  
It also outperforms all unsupervised learning-based methods in Dice (\%).  
However, when trained with label supervision, multi-scale learning-based methods like CorrMLP and RDP surpass all iterative methods in registration accuracy.  
The proposed VoxelOpt achieves Dice (\%) on par with the state-of-the-art (SOTA) learning-based method RDP (semi-supervised), with similar runtime (\(<1\)s), while reducing HD95 by 7.6\%.  
Moreover, VoxelOpt outperforms the best unsupervised learning method CorrMLP by 14.7\% in Dice (\%) and the best iterative method Deeds by 9.2\%, with a substantial runtime reduction.  
This demonstrates that, with readily available foundational models, VoxelOpt offers competitive performance without hand-crafted features, complex contrastive learning, or label supervision.

\begin{table*}[t]
\centering
\caption{
Ablation study of VoxelOpt on feature type, kernel size \(k\), adaptive message passing, and pre-filtering of the cost volume before optimization.
Models are numbered for clarity.
``Foundation'' refers to features extracted using a foundational segmentation model.  
}
\resizebox{1.0\columnwidth}{!}{  
\begin{tabular}{ccccccccc}  
    \hline  
    \hline  
    ~~Model~~ & ~Feature Type~ & ~Kernel Size $k$~ & ~Adaptive~ & ~$\mathbf{C}^k$ Filtering~ & Dice (\%) & HD95 $\downarrow$ & SDLogJ $\downarrow$ & Runtime (s) $\downarrow$ \\  
    \hline  
    Initial & - & - & - & - & 30.86 & 29.77 & - & - \\  
    \hline
    \#1 & Raw Image  & 1 & Yes & Yes & 45.67 & 26.14 & 0.18 & $<1.0$ \\
    \#2 & MIND       & 1 & Yes & Yes & 49.98 & 24.97 & 0.16 & $<1.0$ \\
    \#3 & Foundation & 1 & Yes & Yes & 58.51 & 18.54 & 0.21 & $<1.0$ \\
    \#4 & Foundation & 1 & No  & Yes & 56.40 & 18.85 & 0.20 & $<1.0$ \\
    \#5 & Foundation & 2 & Yes & Yes & 57.95 & 19.21 & 0.25 & 1.0 \\
    \#6 & Foundation & 3 & Yes & Yes & 56.78 & 20.58 & 0.27 & 2.5 \\
    \#7 & Foundation & 1 & Yes & No  & 56.93 & 19.47 & 0.19 & $<1.0$ \\
    \#8 & Foundation & 1 & No  & No  & 54.75 & 19.44 & 0.17 & $<1.0$ \\
    \hline  
    \hline  
\end{tabular}  
\label{tab:ablation}  
}  
\end{table*}

\subsection{Ablation Studies}

In this section, we analyze factors impacting VoxelOpt's performance.  
When varying one factor, others remain fixed at the default settings described in \S\ref{sec:base_imp_eva}.  

\textbf{Effects of Image Features:}  
Table~\ref{tab:ablation} shows that performance improves progressively when using raw images (\#1), MIND features \cite{heinrich2012mind} (\#2), and pre-softmax features from a foundational segmentation model (\#3).  
This aligns with Fig.~\ref{fig:entropy_distribution}, where raw images struggle to provide strong displacement signals, and MIND features introduce local noise in uniform regions.  

\textbf{Effects of The Kernel Size \(k\):}  
Comparing models \#3, \#5, and \#6, increasing the kernel size does not improve registration accuracy but degrades deformation smoothness and exponentially increases computational complexity, as the cost volume space grows from \((2 \cartprod 1 + 1)^3\) to \((2 \cartprod 2 + 1)^3\) and \((3 \cartprod 1 + 1)^3\), with runtime increasing from \(<1.0\)s to 1.0s and 2.5s.  
This demonstrates the superiority of the multi-level image pyramid and 26-neighborhood approach.  

\textbf{Effects of Voxel-Adaptive Message Passing:}  
Comparing models \#3 and \#4, voxel-adaptive message passing improves Dice by 3.7\%.  
While the foundational segmentation model contributes significantly (comparing \#1 and \#3), the entropy measure of displacement signal strength is crucial for identifying pre-softmax segmentation features as naturally suitable for image registration.  

\textbf{Effects of \# of Iterations:}
We vary the number of optimization iterations from 1 to 9 and observe that while Dice slightly increases with more iterations, the change remains $< 0.5\%$, likely because adaptive cost volume filtering already propagates key displacement signals.
As shown in Table 2, comparing \#3 vs. \#7 and \#4 vs. \#8, removing pre-optimization cost volume filtering consistently degrades performance, with or without adaptive filtering.

\section{Discussions \& Conclusions}

ConvexAdam (CA) \cite{siebert2024convexadam} is perhaps the most comparable method to \textbf{VoxelOpt}, but we highlight three key differences that enable a better accuracy-efficiency trade-off:  
1) \textbf{VoxelOpt} consistently uses a fixed 27-neighbor for cost volume construction, avoiding CA’s exponential complexity and enabling full-resolution processing.  
2) \textbf{VoxelOpt} performs one forward pass for displacement extraction, whereas CA involves a two-stage process, discrete followed by iterative, which requires downsampling to remain efficient. 
Using the same MIND feature, our Dice is comparable to CA’s but with significantly lower runtime (7s vs. <1s, see \#2 in Table 2 vs. ``CA'' in Table 1).  
3) \textbf{VoxelOpt} introduces displacement entropy to reveal the effectiveness of different features for registration, facilitating the use of various pre-trained foundational models beyond the one used here.
4) Unrolling optimization into neural networks is common \cite{aggarwal2018modl,zhang2023laro} in medical image reconstruction, but in registration it often incurs high computational cost \cite{heinrich2019closing} or requires precise architectural changes \cite{zhang2025unsupervised}.  
In contrast, \textbf{VoxelOpt} can be seamlessly integrated into networks for end-to-end training.

In conclusion, \textbf{VoxelOpt} bridges the gap between learning-based and iterative methods for deformable image registration through a discrete optimization framework. It achieves state-of-the-art accuracy with sub-second runtime, outperforming iterative methods in efficiency and matching semi-supervised approaches in accuracy, without relying on label supervision or complex training.



\newpage

\bibliographystyle{splncs04}
\bibliography{mybibliography}

\end{document}